\documentclass[twocolumn,tighten]{aastex62}

 \usepackage{booktabs}

\usepackage{bm}
\usepackage{graphicx}

\submitjournal{ApJL}
\shorttitle{Relativistic spin precession in the binary PSR J1141$-$6545}
\shortauthors{Venkatraman Krishnan et al.}
\begin{document}

\title{Relativistic spin precession in the binary PSR J1141$-$6545}

\correspondingauthor{Vivek Venkatraman Krishnan}
\email{vivekvenkris@gmail.com,vkrishnan@swin.edu.au}

\author[0000-0001-9518-9819]{V. Venkatraman Krishnan}
\affil{Centre for Astrophysics and Supercomputing,Swinburne University of Technology, H11, PO Box 218, VIC 3122, Australia.}
\affil{ARC Centre of Excellence for All-sky Astrophysics (CAASTRO)}
\affil{Max-Planck-Institut f{\"u}r Radioastronomie, Auf dem H{\"u}gel 69, 53121 Bonn, Germany }

\author[0000-0003-3294-3081]{M. Bailes}
\affil{Centre for Astrophysics and Supercomputing,Swinburne University of Technology, H11, PO Box 218, VIC 3122, Australia.}
\affil{ARC Centre of Excellence for All-sky Astrophysics (CAASTRO)}
\affil{ARC Centre of Excellence for Gravitational Wave Discovery (OzGrav)}

\author[0000-0003-2519-7375]{W. van Straten}
\affil{Institute for Radio Astronomy and Space Research, Auckland University of Technology}

\author[0000-0002-4553-655X]{E. F. Keane}
\affil{SKA Organisation, Jodrell Bank Observatory, SK11 9DL, UK.}
\affil{Centre for Astrophysics and Supercomputing,Swinburne University of Technology, H11, PO Box 218, VIC 3122, Australia.}

\author[0000-0002-4175-2271]{M. Kramer}
\affil{Max-Planck-Institut f{\"u}r Radioastronomie, Auf dem H{\"u}gel 69, 53121 Bonn, Germany }
\affil{ARC Centre of Excellence for Gravitational Wave Discovery (OzGrav)}

\author[0000-0002-8383-5059]{N.D.R. Bhat}
\affil{International Centre for Radio Astronomy Research, Curtin University, Bentley, WA 6102, Australia.}
\affil{ARC Centre of Excellence for All-sky Astrophysics (CAASTRO)}

\author[0000-0003-1110-0712]{C. Flynn}
\affil{Centre for Astrophysics and Supercomputing,Swinburne University of Technology, H11, PO Box 218, VIC 3122, Australia.}
\affil{ARC Centre of Excellence for All-sky Astrophysics (CAASTRO)}

\author[0000-0003-0289-0732]{S. Os{\l}owski}
\affil{Centre for Astrophysics and Supercomputing,Swinburne University of Technology, H11, PO Box 218, VIC 3122, Australia.}

\begin{abstract}
PSR J1141$-$6545 is a precessing binary pulsar that has the rare potential to reveal the two-dimensional structure of a non-recycled pulsar emission cone. It has undergone $\sim25\degr$ of relativistic spin precession in the $\sim18$ years since its discovery. In this paper, we present a detailed Bayesian analysis of the precessional evolution of the width of the total intensity profile, to understand the changes to the line-of-sight impact angle ($\beta$) of the pulsar using four different physically motivated prior distribution models. Although we cannot statistically differentiate between the models with confidence, the temporal evolution of the linear and circular polarisations strongly argue that our line-of-sight crossed the magnetic pole around MJD 54000 and that only two models remain viable. For both these models, it appears likely that the pulsar will precess out of our line-of-sight in the next $3-5$ years, assuming a simple beam geometry. Marginalising over $\beta$ suggests that the pulsar is a near-orthogonal rotator and provides the first polarization-independent estimate of the scale factor ($\mathbb{A}$) that relates the pulsar beam opening angle ($\rho$) to its rotational period ($P$) as $\rho = \mathbb{A}P^{-0.5}$ : we find it to be $> 6 \rm~deg~s^{0.5}$ at 1.4 GHz with 99\% confidence. If all pulsars emit from opposite poles of a dipolar magnetic field with comparable brightness, we might expect to see evidence of an interpulse arising in PSR J1141$-$6545, unless the emission is patchy.

\end{abstract}
\keywords{stars: neutron --- pulsars: individual (PSR J1141$-$6545) --- relativistic processes  --- radiation mechanisms: non-thermal 
}

\section{Introduction} 
\label{sec:intro}

Binary pulsars with short orbital periods exhibit a wide range of relativistic phenomena \citep{DamourTaylor1992}. These manifest themselves in, for instance, the rate of advance of periastron ($\dot{\omega}$), the amplitude of time dilation ($\gamma$), the time derivative of the orbital period ($\dot{P}_\mathrm{b}$) and the range ($r$) and shape ($s$) of the Shapiro delay. Such effects are usually detected through pulsar timing, a technique where one measures the spin, Keplerian and relativistic dynamics of the pulsar by monitoring the times of arrivals (ToA) of its pulses. The measured relativistic dynamics are usually phenomenologically described by the so-called post-Keplerian formalism \citep{DD1,DD2}, using which predictions of theories of gravity such as the general theory of relativity (GR) may be investigated for consistency. In systems where the spin axis of the pulsar is misaligned with the orbital angular momentum, yet another effect can be potentially observed. Named ``geodetic'' or ``de-Sitter'' precession, this is a relativistic spin-orbit coupling effect where the spin axes of the component stars of a binary system precess around the vector sum of the orbital and spin angular momenta \citep{DamourAndRuffini1974,Barker&Oconnel1975,DamourTaylor1992}. The angular rate of such precession (in $\rm rad$ $\rm s^{-1}$) within GR is given by 
\begin{equation}\label{eq:geod}
\Omega_{\rm geod}  =   {n}^{5/3}  {T_\odot}^{2/3} m_{\rm c} \, \frac { (4m_{\rm p} + 3m_{\rm c})}{2(m_{\rm p} + m_{\rm c})^{4/3}} \frac{1}{1-e^2}
\end{equation}

\noindent where $n = 2\pi/P_{\rm b} $ is the angular velocity of the orbit with period $P_{\rm b}$ in seconds, $T_\odot = GM_\odot/c^3 =$ $4.925490947$ $ \rm \mu s$, $m_{\rm p}$ and $m_{\rm c}$ are the masses of the pulsar and the companion respectively, in units of solar masses ($M_\odot$) and $e$ is the orbital eccentricity \citep{Lorimer&Kramer2005}. Relativistic spin precession changes the viewing angle of the pulsar beam from the Earth, causing secular variations in the observed pulse profile. Such variations have been seen in several relativistic pulsars in the past including the  Hulse-Taylor pulsar PSR B1913+16 \citep{Kramer1998}, PSR B1534+12 \citep{Stairs2004,FonsecaEtAl2014}, the double pulsar PSR J0737$-$3039B \citep{BurgayEtAl2005a,BretonEtAl2008}, PSR J1906+0746 \citep{DavignesEtAl2013}  and PSR J1141$-$6545 \citep{HotanEtAl2004b,ManchesterEtAl2010}.

PSR J1141$-$6545 (hereafter ``the pulsar'') is a young, relativistic binary pulsar in a $\sim4.74$-hr eccentric ($e \sim 0.17$) orbit around a massive ($\sim 1$ $M_\odot$) white-dwarf companion. It was discovered in 2000 in the Parkes Multibeam Pulsar Survey \citep{KaspiEtAl2000} and regular pulsar timing observations have been carried out since then. Given the compact configuration of the binary system, $\dot{\omega}$, $\gamma$ and $\dot{P}_\mathrm{b}$ were soon measured, leading to a test of GR with $\sim 25 \%$ precision \citep{BailesEtAl2003}. \citealt{BhatEtAl2008} performed a $\sim 6 \%$ test of GR along with the estimates of the inclination angle of the system to be $\sim 71 \degr$, whose equally likely degerate solution of $\sim109 \degr$ is now ruled out by a recent study of the annual variations of the pulsar's scintillation velocity (Reardon et al. submitted). The GR masses of the pulsar and the companion were obtained through pulsar timing, providing an estimate of geodetic precession rate of the pulsar of $ 1.36 \degr\rm~yr^{-1}$, implying a precession period of $\sim265$ years (\citealt{HotanEtAl2004b}; hereafter H05). As such a precession rate would imply, the observations also revealed dramatic changes to the pulse profile whose detailed investigations were performed using both the total intensity profile and the polarized emission (\cite{ManchesterEtAl2010}, hereafter M10). 

\begin{figure*}[t]
\label{fig:angles}
\centering 
\includegraphics[scale=0.7, trim=4 4 4 4,clip]{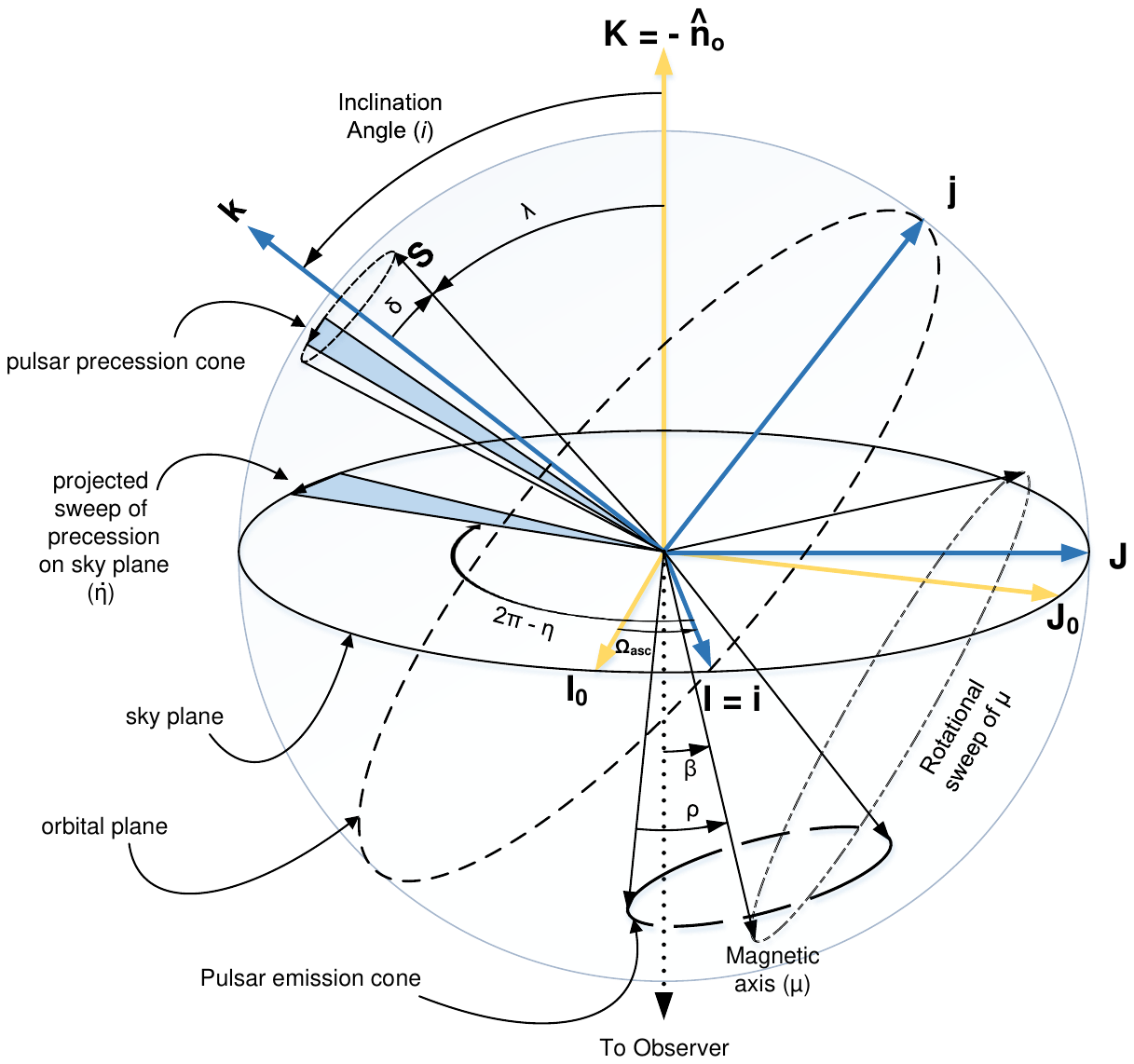} 
\caption{The orientation of the binary system on the sky and the definition of the angles, following the DT92 convention \citep{DamourTaylor1992}. The orbital plane defined by vectors (\textbf{I$\tbond$i}, \textbf{j}) is inclined at an angle \textit{i} to the sky plane defined by vectors \textbf{$ \rm \mathbf{I_0}$} and \textbf{J} and rotated in azimuth by the longitude of the ascending node ($\Omega_{asc}$). The observer's line-of-sight (\textbf{$\mathbf{n_0}$}) is defined as the direction along the negative \textbf{K} axis. The orbital angular momentum vector (\textbf{L}) is by definition along the direction of the unit vector \textbf{k} which is perpendicular to the orbital plane and the spin angular momentum of the pulsar (\textbf{S}) is misaligned from \textbf{L} by the misalignment angle, $\delta$. \textbf{L} and \textbf{S} precess about their vector sum \textbf{J} under GR, sweeping out precessional cones. Given the relative magnitudes of \textbf{S} and \textbf{L}, the precession of \textbf{L} around \textbf{J} is negligible and an assumption $\frac{\mathbf{L}}{|\mathbf{L}|} \simeq \frac{\mathbf{J}}{|\mathbf{J}|} \simeq$ \textbf{k} is usually made. The precessional sweep of \textbf{S}, projected onto the sky plane and measured from \textbf{I} defines the longitude of precession ($\eta$), whose complementary angle is shown. The precession of \textbf{S} also changes the magnetic axis of the pulsar (\textbf{$\mu$}) which changes the impact angle ($\beta$) of our line-of-sight to $\mu$ with time. If $|\beta|$ is less than the opening angle of the pulsar emission cone ($\rho$), we see the pulsar's emission. The angle between the pulsar spin axis and our line-of-sight is $\lambda$ which is equal to $180 \degr -\alpha - \beta$ where $\alpha$ is the angle between \textbf{S} and \textbf{$\mu$}.}
\end{figure*}

The variation of the polarization position angle (P.A.) across the pulse longitude ($\Phi$) is often well described by the ``Rotating Vector Model'' (RVM; \citealt{Radhakrishnan&Cooke}) in which the P.A. ($\Psi$) per pulse longitude follows the relation 

\begin{equation}
\tan(\Psi-\Psi_0) =  \frac{\sin \alpha \sin(\Phi - \Phi_0)}{\sin\zeta \cos \alpha - \cos\zeta \sin \alpha \cos(\Phi - \Phi_0)}
\end{equation}

\noindent where $\alpha$ is the magnetic inclination angle, $\beta$ is the impact angle of our line-of-sight to the magnetic axis ($\mu$), $\zeta = 180 - \lambda = \alpha + \beta$ (see Fig.~\ref{fig:angles} for the angle definitions), and $\Psi_0$ is the central P.A. at the fiducial longitude ($\Phi_0$)\footnote{In this equation and everywhere else in this paper, $\Psi$ is defined following the convention used by \citet[hereafter DT92]{DamourTaylor1992} where the measured P.A. increases \textit{clockwise} on the sky which is opposite to the IAU convention.}. H05 used the steepest section of the P.A. curve, which follows the much simpler relation 

\begin{equation}
\left(\frac{d\Psi}{d\Phi} \right)_{\rm max} =  \frac{\sin \alpha }{\sin \beta},
\end{equation}
to obtain a constraint on the spin misalignment angle ($\delta$) of the pulsar to be between $15\degr < \delta < 30\degr$. M10 used the evolution of the absolute central polarization position angle $\Psi_0$ of the pulsar using the relation
\begin{equation}
\Psi_0 = \Omega_{\rm asc} + \eta
\end{equation}

\noindent where $\Omega_{\rm asc}$ is the longitude of the ascending node and $\eta$, the longitude of precession (\citealt{Kramer&Wex2009}; see Fig.~\ref{fig:angles}). The precessional change in $\Omega_{\rm asc}$ is negligible as the counter precession of the orbit due to the pulsar's spin is very small, given the relative magnitudes of their angular momenta. Hence the change in $\Psi_0$ directly provides the change in $\eta$, from which other angles are computed. M10 predicted that $\beta$ had reached a minimum value and hence, predicted a reversal of the shape variations in the near future (see Fig.~(17) of \cite{ManchesterEtAl2010}). However, our analysis of data that span almost a decade longer does not show any sign of pulse profile symmetry with the earlier data. In this paper, we take an alternative approach to understand the pulsar's precession through robust estimates of its evolving pulse width as detailed below. 

\section{Methods}
\label{sec:methods}

\subsection{Observations} 
\label{sec:observations}
This pulsar has been observed for the past $\sim18$ years using the central beam of the Parkes 20 cm ``multibeam" receiver  \citep{Staveley-SmithEtAl1996} using 6 different backends viz. the Analog Filterbank System (AFB), Caltech Parkes Swinburne Recorder 2 (CPSR2), Parkes Digital Filterbanks (PDFB1, PDFB2, PDFB3) and the CASPER Parkes Swinburne Recorder (CASPSR). \cite{PPTA_DR1} and references therein provide full backend details. For this analysis, we use only the data from backends that recorded full polarisation information, as we use polarisation to distinguish different evolution models, as explained later in Section \ref{sec:results}. The data were integrated to an initial time resolution of 3 minutes and subjected to a median filter to mitigate against any radio frequency interference. Following flux calibration using observations of the Hydra radio galaxy and polarization calibration using the Measurement Equation Template Matching  (METM; \citealt{vanStraten2013}) technique using PSR J0437$-$4715 as the polarized reference source, $\chi^2$ values of the calibration solutions were estimated and only observations with a reduced $\chi^2 < 1.2$ were chosen for further analysis. 

M10 noticed that the rotation measure (RM) of the pulsar shows unphysical variations when computed with just the central region of the profile while the outer wings of the profile had a relatively constant RM (see Fig.~(9) of M10). Our analysis show that such variations continue to date. While M10 chose to use a single value of RM and ignore the central part of the profile for their analysis, we chose to develop an empirical model wherein we obtain the RM for every observation using the \textsc{rmfit} program in \textsc{psrchive} and fit its temporal evolution with a $\rm 4^{th}$ degree polynomial. For every epoch, once the appropriate RM from the model is installed, we sum the data in time and frequency to produce full polarization pulse profiles for every epoch. While M10 measured the width of the pulse at the 50\% level, we estimate the width of the total intensity profile at 10\% of the peak pulsed flux density. Hence, our estimates of the width have bigger uncertainties. We perform the estimation using the {\sc transitions} pulse width estimation algorithm of the {\sc psrstat} program in {\sc psrchive}\footnote{\url{http://psrchive.sourceforge.net/manuals/psrstat/algorithms/width}}. Following \cite{Maciesiak2011}, we multiply this estimate by 1.1  to obtain the pulse-width at the $1\%$ signal-to-noise level which we assume to be the width of the emitted pulse. We use the polarization information to understand and distinguish between several width evolution models as we describe below.

\subsubsection{Width evolution models}
Assuming a circularly symmetric cone of radio emission from the pulsar, one can geometrically relate the instantaneous pulse widths (W) to the opening angle of the emission cone ($\rho$) as

\begin{equation} \label{eq:rw}
\cos \rho = \cos \alpha \cos \zeta + \sin \alpha \sin \zeta \cos (\mathrm{W}/2) 
\end{equation}
\citep{GilEtAl1984}. For non-recycled pulsars, $\rho$ is generally consistent with the relation 
\begin{equation}
 \rho =\mathbb{A}P_{\rm spin}^{-0.5},
\end{equation}

\noindent where $ P_{\rm spin}$ is the spin period of the pulsar and $\mathbb{A}$ is a constant of proportionality (hereafter the ``scale factor") \citep{Lorimer&Kramer2005}. Several empirical estimates of  $\mathbb{A}$ have been made with an ensemble of pulsars with their angles $(\alpha, \lambda)$ estimated from RVM (see \cite{Maciesiak2011} and references therein). With $\rm  P_{spin}$ measured in seconds, and $\rho$ measured in degrees, the value of $\mathbb{A}$ at 1.4 GHz is estimated to lie in the range 4.9 to 6.5$\rm~deg~s^{0.5}$ (eg: \cite{Rankin1990,Biggs1990,KramerEtAl1994,GH96,Maciesiak2011}). The temporal variations in $\rho$ and $\alpha$ are negligible for the timescale of our dataset, so the evolution of W is expected to track $\beta$. 

We performed Markov Chain Monte Carlo (MCMC) fits of Eqn.~(\ref{eq:rw}) to our data. Firstly, we used a mean-shift clustering algorithm to group observations that are closely spaced in time, resulting in $\sim 34$ ``clusters'' ($C_{\rm i}$), each of which is assigned one model parameter $\rm \beta_i$ to denote the impact angle at that time. We then added one global model parameter each for $\alpha$ and $\mathbb{A}$ for a total of 36 parameters. We set uniform priors for $\alpha$ between $0\degr$ and $180\degr$\footnote{ We note that there have been a number of probability distributions discussed in the literature for $\alpha$ (eg: \citep{GilEtAl1996,ZhangEtAl2003}). However, we think it is best to provide here the most conservative estimate of $\alpha$ with a non-informative prior. We also report that changing the prior to the naturally expected distribution of $\sin (\alpha)$ provides posterior distributions consistent with what is presented here.} and $\mathbb{A}$ between $4$ and $8\rm~deg~s^{0.5}$ with the rationale that uniform priors are non-informative, and offer no biases to our posterior estimates. 

Our initial model fits with the prior on all $\rm \beta_i$ as $\rm U(-\rho, \rho)$ resulted in axisymmetric, bi-modal posterior probability distributions for $\beta_{\rm i}$. The two modes of the posterior distributions were non-overlapping for most of the dataset, except for when $\rm \beta_i \to 0$ where the distinct modes merged into a single distribution. To remain agnostic about the sign of $\rm \beta_i$, we used 4 different prior models ($\rm M^{prior}_k \forall  k = \{1,2,3,4\}$),  each with different priors on $\rm \beta_i$ (see Table \ref{tab:beta_models}), thereby breaking the bi-modal posterior degeneracy on $\rm \beta_i$ for every $\rm M^{prior}_k $. The extrema across all the models were chosen to be between $-\rho$ and $\rho$ as $|\beta| < \rho$ is necessary for pulse detection. The models $\rm M^{prior}_1$ and $\rm M^{prior}_2$ assume the signs of $\rm \beta_i$ stay negative and positive respectively for the entire dataset, while $\rm M^{prior}_3$ and $\rm M^{prior}_4$ assume there is a sign flip at MJD 54000. To make sure that the uncertainties on $\rm \beta_i$ are estimated correctly for cases where $\rm \beta_i \to 0$, an additional $\pm 1\degr$ was added to the prior limits whose extremum was otherwise zero. This 1$\degr$ was chosen based on the fact that the average 99\% confidence interval on the estimate of $\rm \beta_i$ was $< 1\degr$. The choice of MJD $=54000$ as the pivotal cluster point that distinguishes the models was motivated by three reasons. Firstly, M10's analysis points to a minimum value of $|\rm \beta_i|$ around this MJD. Secondly, the first indication of a sign flip in the circular polarization profile also happens around this MJD (see Fig.~\ref{fig:pol_evol}). Thirdly, the pulsar experienced a rotational glitch soon after this MJD (at MJD$\sim54272$).

\begin{figure*}[t]
\centering
\begin{tabular}{c}
\includegraphics[scale=0.8]{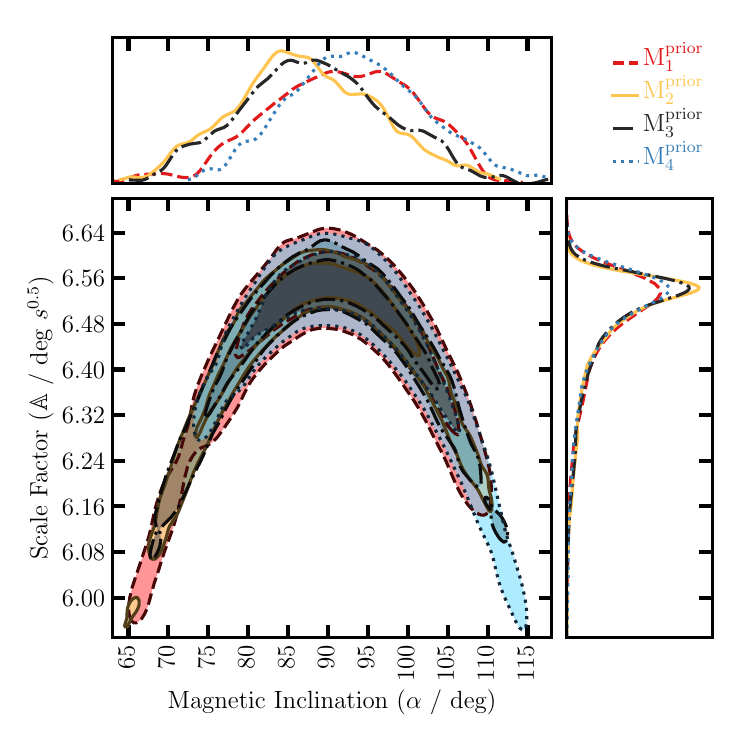} \\ 
\includegraphics[scale=0.8]{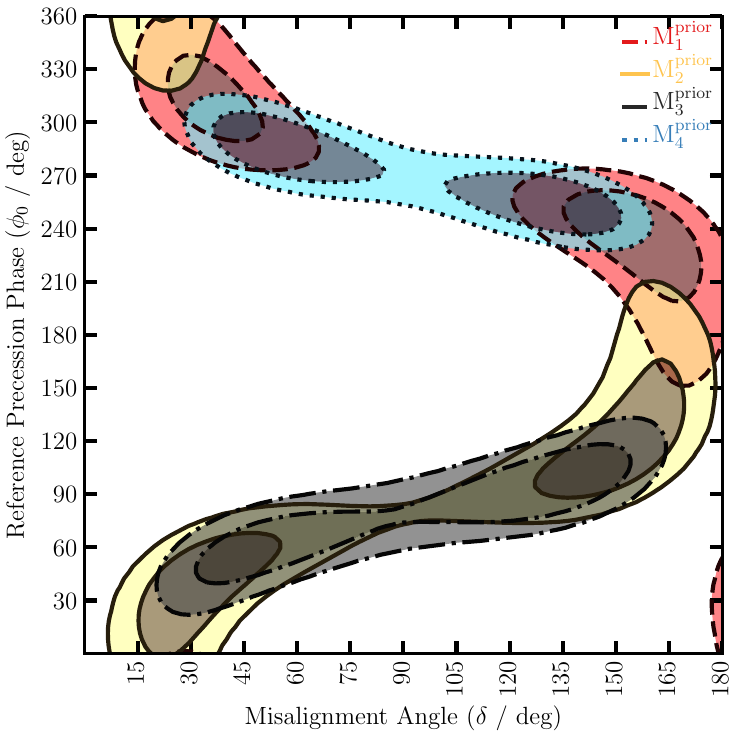} \\
\end{tabular}
\caption{Top: The 68 \% and 95\% contour confidence intervals of the magnetic inclination angle ($\alpha$) and the scale factor ($\mathbb{A}$) along with their marginalised posterior distributions for all $\rm M_k^{prior}$. Bottom: The corresponding contours of the spin-misalignment angle ($\delta$) and the reference precession phase ($\phi_0$).}
\label{fig:posteriors}
\end{figure*}

\begin{table*}[t]
    \centering
    \caption{Model priors for $\rm \beta_i$ with corresponding relative Bayesian Information Criterion values and posteriors for $\alpha$ and $\mathbb{A}$ with $68\%$ confidence intervals.}
    \label{tab:beta_models}
    \begin{tabular}{llccccc}
        \hline
        Model & prior on $\rm \beta_i$ &  $\alpha$ (degrees) & $\mathbb{A}$ & $\Delta \rm BIC$ & $\delta$  (degrees)& $\phi_0$  (degrees)\\
        \hline
        $ \rm M^{prior}_1$ & U($-\rho$, 1)&$90^{+12}_{-9}$ & $6.53^{+0.06}_{-0.10}$ & 0.0& $38 \pm 13$ and $155 \pm 20$&$226 \pm 36$ and $314 \pm 24$ \\ 
        $ \rm M^{prior}_2$ & U(1, $\rho$)&$84^{+13}_{-6}$ & $6.54^{+0.03}_{-0.10}$ & 0.3& $35 \pm 21$ and $149 \pm 21$ & $33 \pm 36$ and $132 \pm 44$\\ 
        $ \rm M^{prior}_3$ & U($-\rho$, 1) MJD $\le 54000$  &$88^{+9}_{-10}$ & $6.54^{+0.04}_{-0.10}$ & 0.1& $91 \pm 60$ & $81 \pm 39$\\ 
        &U($-1$, $\rho$) otherwise  &&&&\\
        $ \rm M^{prior}_4$ & U($-1$, $\rho$) MJD $\le 54000$  &$93^{+9}_{-9~}$ & $6.54^{+0.05}_{-0.10}$ & 0.5& $60 \pm 24$ and $126 \pm 24$&$273 \pm 35$\\ 
        &U($-\rho$, 1) otherwise  &&&&\\
        \hline
    \end{tabular}
\end{table*}

For every $\rm M^{prior}_k$, we marginalize over $\mathbb{A}$ to infer the values of \{$\alpha$, $\rm \beta_i$\}. We used the Gelman-Rubin criterion to assess the convergence of our MCMC chains and used maximum likelihood statistics to compute the parameter uncertainties given the asymmetric posterior distributions (using the \textsc{ChainConsumer} package; \citealt{ChainConsumer}). For each of our MCMC point ($P_{\rm j}$), we obtain $\rm \lambda_i$ from $\alpha$ and $\rm \beta_i$. With this, we perform another MCMC fit to estimate the angles $\phi_0$  and $\delta$ using the relations 

\begin{equation} \label{eq:sg1}
\cos\lambda = \cos i \cos \delta - \sin \delta  \sin i \cos \phi,
\end{equation}

\begin{equation} \label{eq:sg3}
 \phi = \phi_0 +  \Omega_{\rm geod} ( t-t_0),
\end{equation} 
\\
\noindent where $\phi_0$ is the reference precession phase at time $t=t_0$ \citep{DamourTaylor1992}; where $t_0$ is set to MJD~52905. For every MCMC point in the second run ($Q_{\rm k}$), we iterate over each of $P_{\rm j}$, and compute the $\chi^2$ of fitting the function given by Eqn. \ref{eq:sg1} and \ref{eq:sg3} with the values ($\phi_0$, $\delta$) from $Q_{\rm k}$ to $\lambda_{\rm i}$ from $P_{\rm j}$. The likelihood of  $Q_{\rm k}$ is then defined as the sum of the $\chi^2$ over all $P_{\rm j}$. Here we use the inclination angle value of $71 \degr$ obtained from pulsar timing \citep{BhatEtAl2008} and use the GR value for  $\Omega_{\rm geod}$ obtained from Eqn.~\ref{eq:geod}.  
\section{Results and Discussion}
\label{sec:results}
 The posterior distributions of \{$\alpha$, $\mathbb{A}$\} $\forall \rm M^{prior}_k$  after marginalising over $\rm \beta_i$ are shown in Fig.~\ref{fig:posteriors} and their 68\% confidence limits are presented in Table \ref{tab:beta_models}. This analysis provides the first self-consistent estimate of $\mathbb{A}$ independent of the pulsar's polarization profile. As seen in Fig.~\ref{fig:posteriors},  despite being asymmetric with a leading tail, the posterior distribution of $\mathbb{A}$ is confined to be $> 6 \rm~deg~s^{0.5}$ with 99\% confidence. 

Marginalising over $\rm M^{prior}_k$ and $\mathbb{A}$ suggests that the pulsar is a nearly orthogonal rotator with $\alpha = 89\degr^{+18}_{-17}$ at 68\% confidence. Such an orientation, combined with $7\degr$ to $14\degr$ of precession of $\beta$, could have resulted in the detection of the pulsar's interpulse. However, an interpulse has not been observed in our dataset. Given the narrow duty cycle of the pulsar, if one assumes the Double Pole - InterPulse (DP-IP) model of (the lack of) interpulse emission \citep{Lorimer&Kramer2005}, then a further constraint can be added on the posterior distribution of $\alpha$ that 
\begin{equation}
2 \alpha    <  \pi - ( \beta + \rho).
\end{equation}
This constraint rules out the entire 68\% confidence interval on $\alpha$ for every $\rm M^{prior}_k$. Another possibility is that the other pole's emission is fainter than our detection threshold. If so, future observations with the new Parkes Ultra Wideband Low (UWL) receiver \citep{Dunning2015} and the MeerKAT telescopes \citep{BailesEtAl2018}, with their much improved sensitivity and frequency coverage, might be able to detect such an interpulse, which will confirm our estimates of $\alpha$. Yet another possibility is that our initial assumption of a circularly symmetric emission cone is simplistic. Alternative beam shapes such as fan beam models \citep{Dyks2010,WangEtAl2014} have been proposed to explain the complex structures generally seen in the pulse profiles of other pulsars. Investigating such alternate beam shapes is beyond the scope of this paper.  

\begin{figure*}
\centering
\includegraphics[scale=0.73]{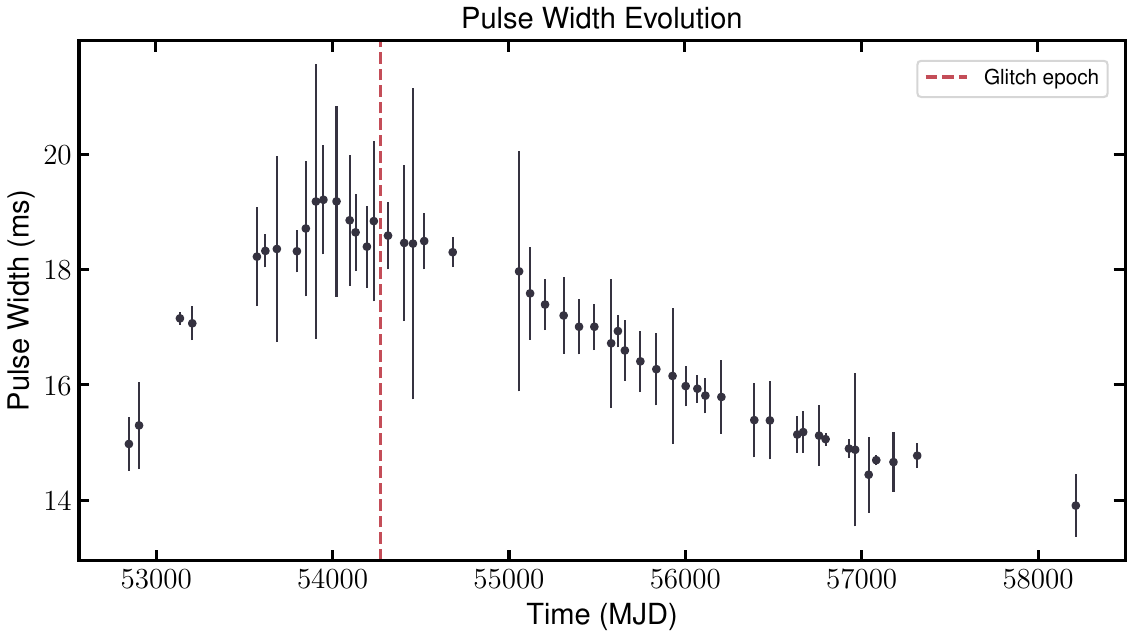} 
\begin{tabular}{cc}
\includegraphics[scale=0.57]{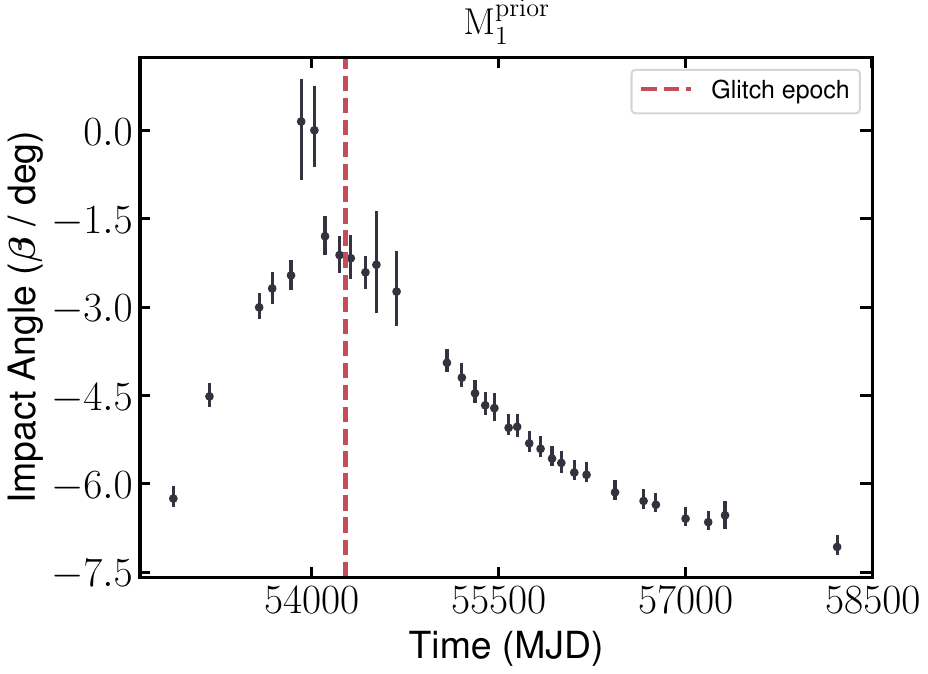} &
\includegraphics[scale=0.57]{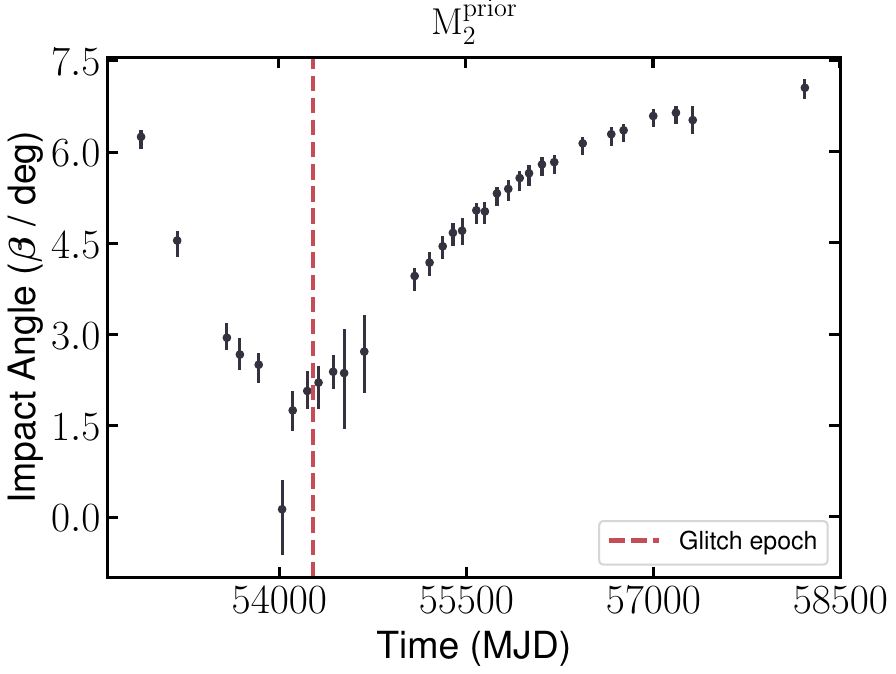} \\
 \includegraphics[scale=0.57]{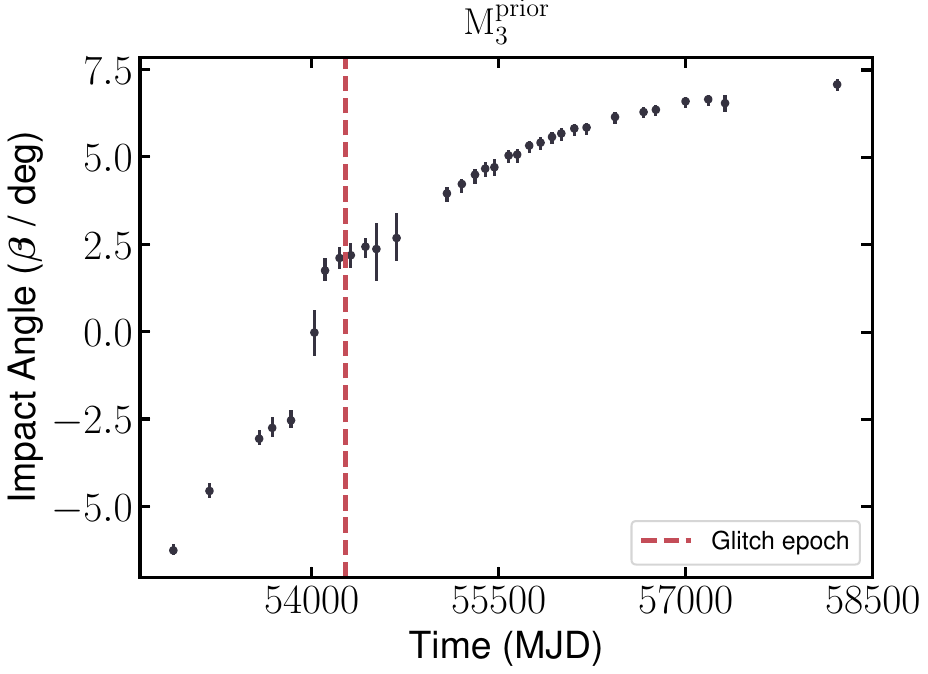} &
\includegraphics[scale=0.57]{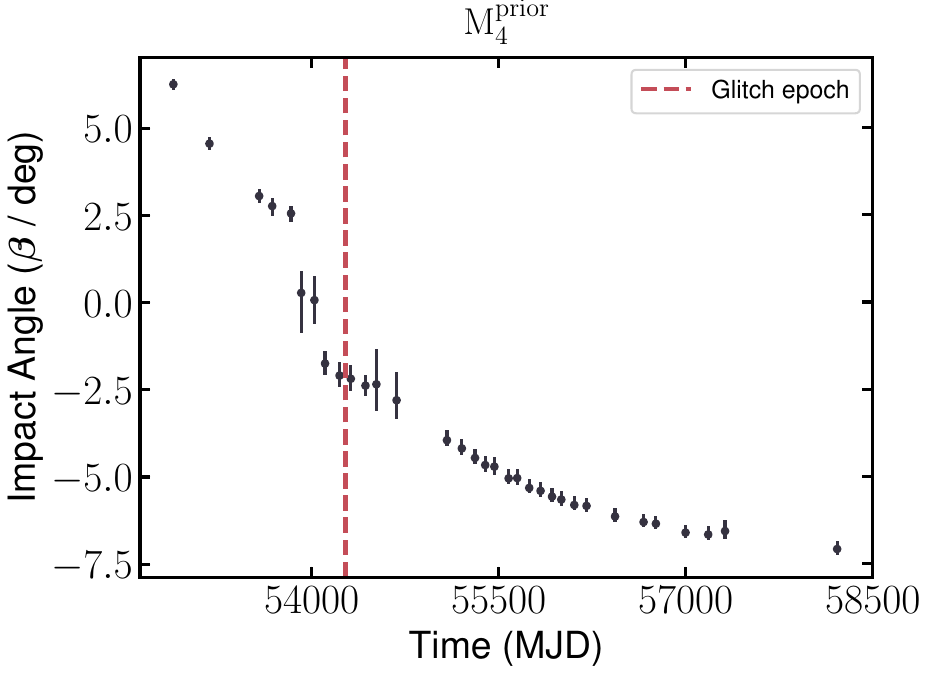} \\
\end{tabular}
\caption{ Top: The temporal evolution of pulse width at the 10\% of the peak pulsed flux density. Remaining panels: The corresponding variations of $\beta$ for the prior models $ \rm M^{prior}_k \forall k = \{1,2,3,4\}.$ The red dashed line in all the plots indicates the glitch epoch. The black dots indicate the mean value and the black lines indicate their corresponding 68\% confidence intervals.}
\label{fig:betas}
\end{figure*}
\begin{figure*}
\centering
\includegraphics[scale=0.5]{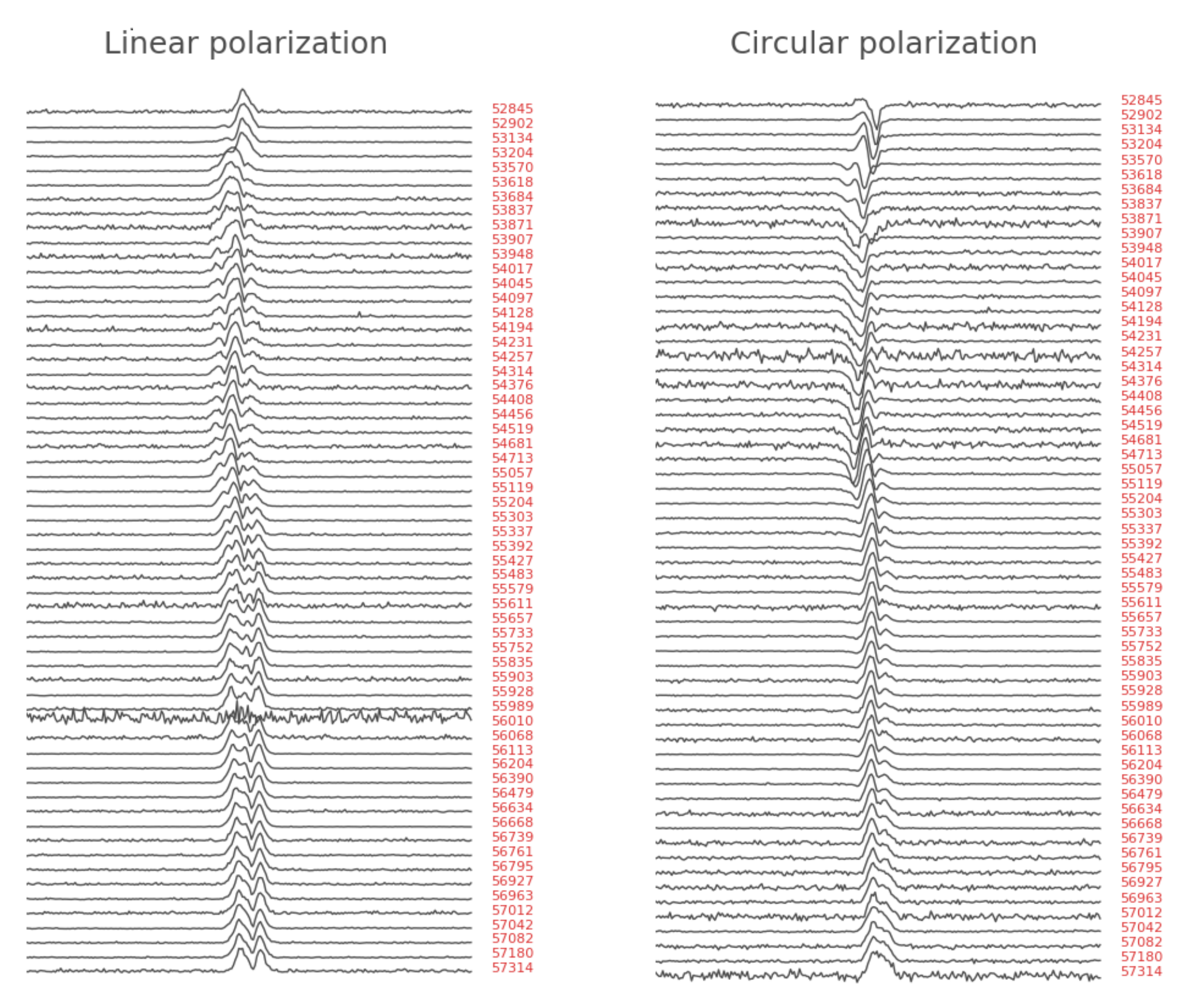} 
\caption{The temporal evolution of linear (left) and circular polarization (right) profiles of PSR J1141$-$6545. Each line spans 0.6 in pulse phase. The number on the right denotes the mean MJD of closely spaced observations clustered using a meanshift estimator (see text). Each line's amplitude is normalised to its peak. The evolution of the polarization rules out any symmetry in the shape variations about MJD $\sim 54000$. The coincident sign flip in circular polarization suggests that the line-of-sight may have crossed the magnetic axis of the pulsar.}  
\label{fig:pol_evol}
\end{figure*}

Our results are in striking contrast with the $1 \sigma$ estimates of $\alpha=160\degr_{-16}^{+8}$ obtained by M10 using the P.A. profile. For conventional models, such a value seems unphysical for a number of reasons. Firstly, assuming M10's value of $\alpha$, one can compute the expected pulse width for every $\rm \beta_i$. Even with a conservative marginalisation over just the unbiased uniform prior probability distribution of $\mathbb{A}$ between $4.9$ and $6.5\rm~deg~s^{0.5}$, the change of the pulse widths for this $\alpha$ is expected to be between $\sim15 \degr$ and $\sim41 \degr$, regardless of the sign of evolution of $\rm \beta_i$. However, as seen in Fig. \ref{fig:betas}, the observed evolution of the pulse width is only between $\sim 5 \degr$ and $\sim16 \degr$. For such values of $\alpha$ to match the observed pulse-widths, the value of $\mathbb{A}$ must be tuned to $\sim 4$. Such a value of $\mathbb{A}$ has not been seen in any young pulsar, assuming circularly symmetric beaming. Secondly, M10 suggested that the evolution of $\beta$ reached its maximum value $\beta_{\rm max} \sim -1 \degr$ near MJD 54000. This prompted them to suggest that there would be a ``reversal'' of shape variations into the next decade as the observer's line-of-sight retraces its path. However, as seen in Fig.~\ref{fig:betas} and \ref{fig:pol_evol} the evolution of width and polarization of the pulse profile are not at all symmetric in our significantly longer dataset.

Such inconsistencies are possibly due to the pulsar's complicated P.A. profile deviating from an ideal RVM sweep. Firstly, the central part of the polarization profile appears to evolve with frequency, part of which is seen to be absorbed into RM estimates leading to unphysical pulse-phase dependent, secular variations of inferred RM. M10 fit for the RVM over just the wings of the profile. However, the centre of the profile can be crucial for values of $\beta$ close to 0, as $\beta \to 0 \Rightarrow$ $\frac{d\Psi}{d\Phi} \to \infty$. Secondly, we see orthogonally polarized modes (OPM; \citealt{Gangadhara1997}) in the P.A. sweep that evolve to non-OPM modes over the dataset. Thirdly, such OPM transition, when occurring at the central part of the profile where the slope of P.A. is the steepest, means that one cannot resolve the degeneracy whether to consider a non-orthogonal step change $\Psi_{\rm step}$ to be either the observed value by itself, the value after an OPM correction ($90 \degr \pm \Psi_{\rm step}$) or the value after a phase unwrap ($180 \degr \pm \Psi_{\rm step}$). Such degeneracies can also affect the absolute central P.A. ($\Psi_0$) that M10 used to compute the longitude of precession ($\eta$). Given such complexities in the P.A. swing, we find the RVM to be too simplistic to be used as is for this pulsar. 

Our Bayesian Information Criterion (BIC) test between the 4 models could not clearly distinguish the best model (see Table~\ref{tab:beta_models}). However, there are two physical arguments that could be used to differentiate the models. Firstly, most regions of the posterior distribution of $\phi_0$ for $\rm M^{prior}_1$ and $\rm M^{prior}_2$ fail to predict the sharp turnover (which happen when $\phi = 0 \degr$ or $180 \degr$) in the evolution of $\beta$ seen in these models and hence those models are disfavored. Additionally, the detection of a sign flip in circular polarization around the epoch of minimum $|\beta|$ suggests that our line-of-sight has crossed the magnetic axis during the course of our observing campaign thereby favoring models $\rm M^{prior}_3$ and $\rm M^{prior}_4$. This might explain the fact that we do not see a reversal of shape variations as M10 predicted. If true, regular observations of this pulsar until it precesses out of our line-of-sight will give us the first glimpse of the 2-dimensional structure of a large fraction of the pulsar emission cone. Regardless of the choice of $\rm M^{prior}_k$, our analysis indicates that the pulsar will precess out of our line-of-sight in the next $3-5$ years. The posterior distributions of $\phi_0$ and $\delta$ are plotted in Fig.~\ref{fig:posteriors} and their 68\% confidence intervals are reported in Table \ref{tab:beta_models}. Without knowledge of the evolution of $\beta$, it is presently not possible to significantly constrain the possible values for $\phi_0$ and $\delta$. Future observations with the Parkes UWL receiver might help resolve the ambiguities in the RM of the pulsar which can then be utilised to obtain reliable constraints of the pulsar geometry from its RVM. Comparing such constraints with the ones obtained in this paper might provide further insights on the system's orbital geometry.

It is also interesting that a rotational glitch takes place soon after the supposed reversal. It is possible that M10's projections were correct but that the glitch reconfigured the pulsar's magnetosphere resulting in changes to the observed pulse profile. To check if the glitch had altered $\alpha$, we performed a BIC test of all $\rm M^{prior}_k$ with two model parameters for $\alpha$ at either side of the glitch epoch. This returned consistent posteriors and disfavoured the split of $\alpha$, thus ruling out any major magnetospheric reconfiguration as a result of the glitch. However, we cannot rule out any glitch induced change in emission properties that did not change $\alpha$.
\section{Conclusions}

We performed an analysis of the evolving pulse widths of PSR J1141$-$6545 due to spin precession using $\sim 18$ years of observations with robust polarization calibration and pulse width estimation methods. While we cannot uniquely infer the sign of the impact angle $\beta$ for every observation, the absolute magnitude is well constrained. The circular polarization sign flip at MJD$\sim54000$ combined with the temporally asymmetric shape variations supports a magnetic axis cross-over. Our estimate of the magnetic inclination angle $\alpha$, regardless of $\rm M_k^{prior}$, indicates that the pulsar is a near-orthogonal rotator. The absence of an observed interpulse emission motivates continued monitoring of this pulsar with more sensitive instruments like the Parkes UWL receiver and the MeerKAT telescope.

\label{sec:conclusion}

\acknowledgements

We thank the anonymous referee for a thorough cross-check of our analysis, along with numerous suggestions that have significantly improved the manuscript. We thank N. Wex and P. C. C. Freire for useful discussions and suggestions. The Parkes radio telescope is funded by the Commonwealth of Australia for operation as a National Facility managed by CSIRO. This research was primarily supported by the Australian Research Council Centre of Excellence for All-sky Astrophysics (CAASTRO; project number CE110001020). Computations were performed on the gSTAR/ozSTAR national facilities at Swinburne University of Technology funded by Swinburne and the Australian Government's Education Investment Fund. SO acknowledges the Australian Research Council grant Laureate Fellowship FL150100148. NDRB acknowledges the support from a Curtin Research Fellowship (CRF12228). 


\end{document}